\documentclass[%
reprint,
superscriptaddress,
 amsmath,amssymb,
 aps,
prb,
nofootinbib
]{revtex4-1}

\usepackage{physics}
\usepackage[dvipsnames]{xcolor} 
\usepackage{graphicx}
\usepackage{dcolumn}
\usepackage{bm}
\usepackage{xcolor, soul}
\sethlcolor{yellow}
\usepackage[makeroom]{cancel}

\makeatletter
\def\@fnsymbol#1{%
   \ifcase#1\or
   \TextOrMath \textdagger \dagger\or
   \TextOrMath \textasteriskcentered *\or
   \TextOrMath \textdaggerdbl \ddagger \or
   \TextOrMath \textsection  \mathsection\or
   \TextOrMath \textparagraph \mathparagraph\or
   \TextOrMath \textbardbl \|\or
   \TextOrMath {\textdagger\textdagger}{\dagger\dagger}\or
   \TextOrMath {\textdaggerdbl\textdaggerdbl}{\ddagger\ddagger}\else
   \@ctrerr \fi
}
\makeatother

\usepackage[T1]{fontenc}

\begin{document}
\title{Non-adiabaticity from first principles: the exact-factorization approach for solids}

\author{Galit Cohen}
\thanks{These authors contributed equally to this work.}
\affiliation{Department of Molecular Chemistry and Materials Science, Weizmann Institute of Science, Rehovot 7610001, Israel}

\author{Rachel Steinitz-Eliyahu}
\thanks{These authors contributed equally to this work.}
\affiliation{Department of Molecular Chemistry and Materials Science, Weizmann Institute of Science, Rehovot 7610001, Israel}

\author{E. K. U. Gross}
\thanks{These authors share equal correspondence.}
\affiliation{Fritz Haber Center for Molecular Dynamics, Institute of Chemistry, The Hebrew University of Jerusalem, Jerusalem 91904, Israel}

\author{Sivan Refaely-Abramson}
\thanks{These authors share equal correspondence.}
\affiliation{Department of Molecular Chemistry and Materials Science, Weizmann Institute of Science, Rehovot 7610001, Israel}

\author{Ryan Requist}
\thanks{These authors share equal correspondence.}
\affiliation{Division of Theoretical Physics, Ru\dj er Bo\v{s}kovi\'{c} Institute, Zagreb 10000, Croatia}
\affiliation{Fritz Haber Center for Molecular Dynamics, Institute of Chemistry, The Hebrew University of Jerusalem, Jerusalem 91904, Israel}

\begin{abstract}
The thorough treatment
of electron-lattice interactions from first principles is one of the main goals in condensed matter physics.
While the commonly applied adiabatic Born-Oppenheimer approximation is sufficient for describing many physical phenomena, it is limited in its ability to capture meaningful features originating from non-adiabatic coupling effects. 
The exact factorization method, starting from the full Hamiltonian of electrons and nuclei, provides a way to systematically account for non-adiabatic effects.
This formalism was recently developed into an \textit{ab initio} density functional theory framework.
Within this framework we here develop a perturbative approach to the electronic states in solid state materials. We derive exact-factorization-based perturbations of the Kohn-Sham states up to second order in the nuclear displacements. These non-adiabatic features in the calculated energy and wavefunction corrections are expressed in terms of readily available
density functional perturbation theory components.
\end{abstract}
  
\maketitle

\section{Introduction}
Theoretical understanding of electron-phonon interactions is central to the realization and optimization of  numerous physical phenomena, from the computation of fundamental ground and excited state properties in materials to optoelectronic phenomena and
dynamical scattering processes dominating energy conversion and transport in materials~\cite{giustino2010electron,giustino2008small,eiguren2003electron,valla1999many,lafuente2022ab,lafuente2022unified,sio2023polarons,perfetto2023real,stefanucci2023and}. Calculations of these interactions are typically made within the adiabatic Born-Oppenheimer (BO) approximation~\cite{BO,ziman1960electrons}, in which it is assumed that the electrons adapt 
immediately to the nuclear configuration. As a result of this approximation, the wavefunction describing the system can be expressed as a single product of an electronic part and a nuclear part, where the electronic part depends parametrically on the nuclear positions. 
For many physical phenomena, this separation of the wavefunctions is extremely successful. 
However, it does not capture non-adiabatic (NA) effects~\cite{pisana2007breakdown,malhado2014non,yarkony2012nonadiabatic,yonehara2012fundamental,miglio2020predominance}.

Non-adiabatic effects are observed in the electronic degrees of freedom~\cite{ponce2015temperature,allen2017low,miglio2020predominance,giustino2017electron}, e.g. in energy spectra, kinks in photoemission, charge transfer and localization, polaron formation~\cite{lafuente2022ab,lafuente2022unified,sio2023polarons,mozyrsky2006intermittent,franchini2021polarons},  superconductivity~\cite{novko2016surface,hu2022nonadiabatic} and lattice-induced decoherence~\cite{vindel2021study}, as well as in the phononic ones, for example in frequency renormalization, Raman measurements, and x-ray spectra~\cite{pisana2007breakdown,girotto2023dynamical,girotto2023ultrafast}. 
Accounting for the electron-phonon interaction in materials in a rigorous way, including non-adiabatic effects, is thus substantial for establishing a general theoretical framework. Yet, a full \textit{ab initio} treatment of these phenomena remains a challenging task.

The traditional and most direct way of treating electron-phonon interactions starts from the BO approximation.
An \textit{ab initio} treatment of electrons and phonons within the BO approximation 
is achieved through density functional perturbation theory (DFPT)~\cite{gonze1995adiabatic,baroni2001phonons,giustino2017electron}, by calculating the adiabatic response of the Kohn-Sham potential and BO potential energy surface to small displacements in the nuclear coordinates. It has been proposed~\cite{calandra2010adiabatic} to extend DFPT by accounting for the electronic response to time-dependent nuclear displacements. A common starting point in many approaches to the calculation of electron-phonon interaction is the Fr\"{o}hlich Hamiltonian, which comprises mean-field electrons, BO phonons and the lowest order electron-phonon interaction. The key challenge is deducing a Fr\"ohlich type Hamiltonian from the full \textit{ab initio} Hamiltonian of electrons and nuclei. Several attempts have been made to overcome this difficulty ~\cite{van2004first, Marini2015}.
A closed set of equations, called the Hedin-Baym equations, for electronic and phononic Green's functions has been formulated by Guistino ~\cite{giustino2017electron} starting from the seminal works of Baym and Hedin ~\cite{baym1961field,hedin1965new}. These equations have been further formulated by Stefanucci, van Leeuwen, and Perfetto for out-of-equilibrium systems, in terms of conserving diagrammatic expansions \cite{stefanucci2023and}. 

The work presented in this paper also starts from the full \textit{ab initio} Hamiltonian. It then employs the exact factorization (EF)~\cite{abedi2010exact,gidopoulos2014electronic} of the many-electron many-nucleus wavefunction, which renders the formalism technically similar to the BO$+$DFPT treatment, but it remains formally exact, i.e.~retains all non-adiabatic effects. 
This formalism was employed to deduce a density functional framework for the complete system of electrons and nuclei \cite{requist2016exact, li2018density}. For the case of small amplitude nuclear motion, this can be cast into a density functional theory (DFT) for electrons and phonons\cite{RPG2019}. In this approach the Kohn-Sham electrons are ``dressed'' by
non-adiabatic phonon-induced interactions that appear through a non-adiabatic correction to the Kohn-Sham potential. The self-consistent density obtained is the conditional probability of finding an electron at position $\mathbf{r}$ given that the collection of nuclei resides at a certain point in nuclear configuration space.
As a consequence, one can avoid transformation to the body-fixed coordinate  frame~\cite{van2004first,Kreibich2001}, which is a big advantage in practical terms. 
In Ref.~\onlinecite{RPG2019}, the density functional formalism was applied to a simple Fr\"ohlich model system as a test. 
So far this formalism has not been implemented to treat real materials. We present here a full \textit{ab initio} implementation for extended systems including multiband transitions and higher-order electron-phonon interactions. We derive a theoretical framework within EF-DFT to account for non-adiabatic effects arising from these interactions. 

Our manuscript is organized as follows. We present the general  EF-DFT theory for extended systems in section II. We derive nonadiabatic DFPT-based corrections to the electronic energy bands and wavefunctions in section III, and discuss their implications in section IV.    

\section{Exact factorization approach to non-adiabatic effects in electron-lattice interactions}

We begin with a general description of the full system via the many-body (MB) Hamiltonian
\begin{equation}\label{eq:MB_H}
    \hat{\mathcal{H}} = \hat{T}_n(\mathbf{R})+\hat{W}_{nn}(\mathbf{R})+\hat{T}_e(\mathbf{r})+\hat{W}_{ee}(\mathbf{r})+\hat{W}_{en}(\mathbf{R},\mathbf{r})\;{,}
\end{equation}
consisting of the kinetic and Coulomb interactions in the nuclear and electronic systems and the mutual interaction between electrons and nuclei. The sets of nuclear and electronic coordinates are denoted by $\mathbf{R}$ and $\mathbf{r}$, respectively. The complete many-electron many-nucleus wavefunction satisfies the time-independent Schr\"{o}dinger equation (SE)
\begin{equation} \label{eq:SE}
\hat{\mathcal{H}}\Psi(\mathbf{r},\mathbf{R})=E\Psi(\mathbf{r},\mathbf{R})\;,
\end{equation}
where the eigenstates of the full Hamiltonian, the MB wavefunctions, are denoted by $\Psi(\mathbf{r},\mathbf{R})$~\footnote{While the full Hamiltonian $\hat{\mathcal{H}}$ is translationally invariant, we assume eigenstates with spontaneously broken translational symmetry~\cite{anderson1972more}.}.

Owing to the small ratio of electronic over nuclear mass, one intuitively expects that the electrons quickly adjust to the slow motion of the nuclear degrees of freedom. This intuitive picture suggests the \textit{adiabatic approximation} for the electron-nucleus wavefunction
\begin{equation} \label{eq:psi_adiab}
    \Psi^{ad}(\mathbf{r},\mathbf{R}) = \phi_\alpha^{BO}(\mathbf{r}|\mathbf{R})\chi_{\alpha\lambda}(\mathbf{R})\;.
\end{equation}
Thus, the MB wavefunction is approximated as a single product of separate electronic and nuclear wavefunctions;
$\phi_\alpha^{BO}$ is the electronic wavefunction in the Born- Oppenheimer~\cite{BO} approximation and $\chi_{\alpha\lambda}(\mathbf{R})$ is the nuclear wavefunction.
For each fixed nuclear configuration, the electronic BO wavefunction satisfies 
\begin{equation} \label{eq:SE-el}
    \hat{H}^{BO}(\mathbf{r},\mathbf{R})\phi^{BO}_\alpha(\mathbf{r}|\mathbf{R}) = \varepsilon_\alpha(\mathbf{R})\phi^{BO}_\alpha(\mathbf{r}|\mathbf{R})
\end{equation}
with
\begin{equation}
    \hat{H}^{BO}(\mathbf{r},\mathbf{R}) = \hat{W}_{nn}(\mathbf{R})+\hat{T}_e(\mathbf{r})+\hat{W}_{ee}(\mathbf{r})+\hat{W}_{en}(\mathbf{R},\mathbf{r})\;
\end{equation}
being the Hamiltonian governing the correlated motion of electrons in the field of clamped nuclei.
The nuclear wavefunctions $\chi_{\alpha\lambda}(\mathbf{R})$ are the solutions of the nuclear SE
\begin{equation}  \label{eq:SE-nuc}
    \left(\hat{T}_n(\mathbf{R})+ \varepsilon_\alpha(\mathbf{R})\right)\chi_{\alpha\lambda}(\mathbf{R}) = \Omega_{\alpha\lambda}\chi_{\alpha\lambda}(\mathbf{R})\;,
\end{equation}
where the $\lambda$ index denotes the vibrational states associated with each electronic eigenstate, $\alpha$. The crucial simplification achieved by the adiabatic approximation is that nuclear and electronic degrees of freedom are treated in separate equations, Eqs.~\eqref{eq:SE-el} and \eqref{eq:SE-nuc}, where the solution of Eq.~\eqref{eq:SE-el}, namely the electronic eigenvalue $\varepsilon_\alpha(\mathbf{R})$, is used as input to the nuclear SE, Eq.~\eqref{eq:SE-nuc}. 
The nuclear kinetic energy reads
\begin{equation}
    \hat{T}_n(\mathbf{R})= -\mu\cdot \sum_\kappa \frac{1}{2M_\kappa}\nabla_{R_\kappa}^2
\end{equation}
where $\mu=\frac{m}{M}$ with $m$ being the electron mass and $M$ representing a nuclear reference mass (typically chosen to be the proton mass) such that $M_\kappa\cdot M$ are the true nuclear masses. Here and in the remainder of the article, all equations are written in atomic units ($e^2=\hbar=m=1$).

While in Eq.~\eqref{eq:psi_adiab} the nuclear wavefunction is the solution of Eq.~\eqref{eq:SE-nuc}, one may improve upon Eq.~\eqref{eq:psi_adiab}  by variationally optimizing the nuclear part. This leads to  

\begin{equation}\label{eq:SE_BH}
\begin{split}
    \Big[\mu \sum_\kappa\frac{\left(-i\nabla_{R_\kappa}+A_{\kappa}[\phi^{BO}_\alpha](\mathbf{R})\right)^2}{2M_\kappa}&+\varepsilon_\alpha(\mathbf{R})+\varepsilon^{geo}[\phi^{BO}_\alpha](\mathbf{R}) \Big]\\&\times\Tilde{\chi}_{\alpha\lambda}(\mathbf{R}) =\Tilde{\Omega}_{\alpha\lambda}\Tilde{\chi}_{\alpha\lambda}(\mathbf{R})\;{,}
\end{split}
\end{equation} \\
where $\Tilde{\chi}$ is the nuclear wavefunction that minimizes the total energy.

In Eq.~\eqref{eq:SE_BH} the vector potential and geometric term are defined in terms of the general expressions
\begin{equation} \label{eq:vector_pot}
   A_{\kappa}[\phi_\alpha](\mathbf{R}) = \mel{\phi_\alpha}{-i\nabla_{R_\kappa}}{\phi_\alpha} 
\end{equation} 
\begin{equation} \label{eq:eps_geo}
\begin{split}
    &\varepsilon^{geo}[\phi_\alpha](\mathbf{R}) = \\
    &\;\;\;\mu\sum_\kappa\frac{1}{2M_\kappa}\left[\braket{\nabla_{R_\kappa}\phi_\alpha}{\nabla_{R_\kappa}\phi_\alpha} - A_{\kappa}[\phi_\alpha](\mathbf{R})^2\right]
\end{split}
\end{equation}
evaluated for $\phi_\alpha=\phi_\alpha^{BO}$. Note that $\varepsilon^{geo}$ carries a prefactor of the mass ratio.

Recently, the exact factorization (EF) approach was introduced as a way to include non-adiabatic effects beyond the BO approximation~\cite{abedi2010exact,gidopoulos2014electronic,hunter1975conditional}. This is achieved by expressing the MB wavefunction as a single product of electronic and nuclear exact wavefunctions
\begin{equation}\label{eq:Psi_EF}
    \Psi(\mathbf{r},\mathbf{R})=\phi(\mathbf{r}|\mathbf{R})\chi(\mathbf{R})\;.
\end{equation}
Note that this formally exact representation of the full wavefunction involves an electronic state which is different from the BO electronic wavefunction but shares with it the normalization condition $\int\abs{\phi(\mathbf{r}|\mathbf{R})}^2 dr =1$.
This differs from the Born-Huang expansion~\cite{bornHuang}, in which multiple BO surfaces are directly accounted for, and from the adiabatic approximation in which a single BO surface is considered. Within the EF scheme, non-adiabatic effects are accounted for in a SE equivalent to Eq.~\eqref{eq:SE_BH}, where the vector potential,~Eq.~\eqref{eq:vector_pot}, and the geometric term,~Eq.~\eqref{eq:eps_geo}, are functionals of the exact electronic wavefunction (rather than the BO wavefunation). The equations of motion of the two factors, $\phi(\mathbf{r}|\mathbf{R})$ and $\chi(\mathbf{R})$, can be deduced from the Rayleigh-Ritz variational principle by minimizing the total energy functional 
\begin{equation}\label{eq:totalE}
    \begin{split}
        E=&\mel{\Psi}{\hat{\mathcal{H}}}{\Psi}\\
        =& \int\chi^*(\mathbf{R})\mu\sum_\kappa\frac{1}{2M_\kappa}\left(-i\nabla_{R_\kappa}+A_\kappa[\phi](\mathbf{R})\right)^2\chi(\mathbf{R}) \\ &+ \int\abs{\chi(\mathbf{R})}^2 \left(\varepsilon^{BO}[\phi](\mathbf{R})+\varepsilon^{geo}[\phi](\mathbf{R})\right)
    \end{split}
\end{equation}

with 
\begin{equation}\label{eq:eps_BO}
    \varepsilon^{BO}[\phi](\mathbf{R}) = \int dr \phi^*(\mathbf{r}|\mathbf{R})\hat{H}^{BO}(\mathbf{r},\mathbf{R})\phi(\mathbf{r}|\mathbf{R})\;.
\end{equation}
Using this variational principle, a density functional theory has been formulated in terms of the electronic conditional probability density, $n(\mathbf{r}|\mathbf{R})$~\cite{requist2016exact}. This leads to a Kohn-Sham-like equation for the electronic degrees of freedom, while the nuclear degrees of freedom are described by the N-body SE 
\begin{equation}\label{eq:nuc_eof}
\begin{split}
    \bigg[\mu\sum_\kappa \frac{1}{2M_\kappa}(-i\nabla_{R_\kappa}+A_\kappa[\phi]&(\mathbf{R}))^2
    +\varepsilon^{BO}[\phi](\mathbf{R})\\&+\varepsilon^{geo}[\phi](\mathbf{R})\bigg]\chi(\mathbf{R})=E\chi(\mathbf{R})
\end{split}
\end{equation}
that follows directly from the EF. 

Hereafter, the vector potential is neglected, as justified for systems with vanishing nuclear current density in the ground state. 
Expanding $\varepsilon^{BO}(\mathbf{R})$ and $\varepsilon^{geo}(\mathbf{R})$ to second order in the nuclear displacements from their equilibrium positions allows us to solve Eq.~\eqref{eq:nuc_eof} in terms of harmonic phonon coordinates $U=\{U_{\mathbf{q}\lambda}\}$, with $\mathbf{q}$ and $\lambda$ being the phonon momentum and mode, respectively: 
\begin{equation}\label{eq:EF_nuc_eq}
  \bigg[\mu\sum_{\mathbf{q}\lambda} \frac{\hat{P}_{-\mathbf{q}\lambda}\hat{P}_{\mathbf{q}\lambda}}{2}+ \varepsilon(U)\bigg]\chi(U) = \Omega\chi(U)\;.
\end{equation}

The electronic Kohn-Sham equation reads
\begin{equation}\label{eq:EF_elec_eq}
    \begin{split}
        \Big[\frac{\hat{p}^2}{2m}+\hat{v}_{en}(\mathbf{r},U)+\hat{v}_{hxc}(\mathbf{r},U)+\hat{v}_{geo}&(\mathbf{r},U)\Big] \psi_{n\mathbf{k}U}(\mathbf{r})\\
        &=\epsilon_{n\mathbf{k}U}\psi_{n\mathbf{k}U}(\mathbf{r})\;.
    \end{split}
\end{equation}

Eq.~\eqref{eq:EF_nuc_eq} accounts for the nuclear interactions, where the first term is the nuclear kinetic energy, $\varepsilon(U)$ is the exact potential energy surface, and $\Omega=E-\varepsilon(\mathbf{R_0})$ is the total energy of the phonons. $\mathbf{R_0}$ is the equilibrium position corresponding to the bottom of the BO potential energy surface. 
The EF energy surface $\varepsilon(U)$ is given by
\begin{equation}\label{eq:epsU}
    \varepsilon(U)=\varepsilon^{BO}(U)+\varepsilon^{geo}(U)\;,
\end{equation} 
where we emphasize that both $\varepsilon^{BO}[\phi]$ and $\varepsilon^{geo}[\phi]$ in Eqs.~\eqref{eq:eps_BO} and \eqref{eq:eps_geo} are evaluated with the EF electronic wavefunction and therefore contain all powers of $\mu$.  In the present work, we approximate each term by its lowest order with respect to powers of $\mu$. This corresponds to treating the $\varepsilon^{BO}[\phi]$ term by a standard BO-based DFT functional. For the $\varepsilon^{geo}$ term, which introduces the non-adiabatic ``dressing'' of the electron-phonon interaction, we adopt a determinantal approximation for $\phi$.

Returning to Eq.~\eqref{eq:EF_elec_eq}, we observe that it has the same form as the standard Kohn-Sham equation, where
$\hat{p}^2/2m$ is the electronic kinetic energy, $\hat{v}_{en}$ is the electron-lattice interaction potential, and $\hat{v}_{hxc}$ is the Hartree-exchange-correlation (HXC) potential. Non-adiabaticity is introduced through an additive potential, $\hat{v}_{geo}$, derived from the functional derivative of the $\varepsilon^{geo}$ term in the total energy, Eq.~\eqref{eq:totalE}. In the original EF-DFT~\cite{requist2016exact} framework $\hat{v}_{geo}$, like $\hat{v}_{hxc}$, is a local multiplicative potential. Since in the present work we make the determinantal approximation for $\phi$, $\hat{v}_{geo}$ acts as a non-local potential, in the spirit of the generalized Kohn-Sham (GKS) approach~\cite{kummel2008orbital}, and is a functional of conditional single particle orbitals $\psi_{n\mathbf{k}U}(\mathbf{r},\mathbf{R})$;
$n,\mathbf{k}$ represent an electronic state with band index $n$ and quasimomentum $\mathbf{k}$. Hereafter, the subscript $U$ on the energies and orbitals is suppressed.  Making the orbital-dependent energy functional stationary with respect to variations of the GKS orbitals, subject to orthonormality constraints imposed through a matrix of Lagrange multipliers, gives the matrix elements of $\hat{v}_{geo}$~\cite{RPG2019} as
\begin{equation}\label{eq:v_geo_general}
\begin{split}
    &\bra{\psi_{m,\mathbf{k}+\mathbf{q}}}
    \hat{v}_{geo}\ket{\psi_{n\mathbf{k}}} = \braket{\psi_{m,\mathbf{k}+\mathbf{q}}}{\frac{\delta \varepsilon^{geo}}{\delta\psi_{n\mathbf{k}}^*}}\\
    &-\frac{1}{\abs{\chi}^2}\sum_{\mathbf{q}\lambda}\braket{\psi_{m,\mathbf{k}+\mathbf{q}}}{\frac{\partial}{\partial U^*_{\mathbf{q}\lambda}}\Bigg[\abs{\chi}^2\frac{\delta\varepsilon^{geo}}{\delta(\partial \psi_{n\mathbf{k}}^*/\partial U^*_{\mathbf{q}\lambda})}\Bigg]}\;.
\end{split}
\end{equation}
Alternatively, the orbitals could be restricted to come from a local potential so that minimizing the total energy with respect to this local potential will lead to an OEP-like approach~\cite{gidopoulos2014electronic}.

It is useful to quantify non-adiabatic effects in terms of powers of $\mu$. 
The potential energy surface, being treated as a density functional, generates the GKS potential through functional differentiation.
We see in Eqs.~\eqref{eq:epsU}, \eqref{eq:eps_geo}, and \eqref{eq:eps_BO} that the potential energy surface contains a zeroth-order term $\varepsilon^{BO}$ that has no explicit $\mu$ dependence and a first-order term $\varepsilon^{geo}$ that has a prefactor of $\mu$. As a consequence, the GKS potential in Eq.~\eqref{eq:EF_elec_eq} inherits zeroth-order terms $\hat{v}_{en}$ and $\hat{v}_{hxc}$, and a first-order term given by $\hat{v}_{geo}$. To account for the mutual influence of electrons and phonons, in principle this set of equations must be solved self-consistently. As a first step, in the next section we adapt DFPT to the EF equations.

\section{Non-adiabatic perturbative treatment of electron-phonon interactions}

In order to apply EF-DFT in realistic materials, we now turn to the derivation of the additive potential $\hat{v}_{geo}$. Here, we present a perturbative approach, in which corrections to the KS wavefunctions and the energy surface are introduced within first- and second-order perturbation theory.

The single-particle Hamiltonian can be written as the sum of a non-interacting part and an interaction term,
$\hat{\mathcal{H}}=\hat{\mathcal{H}}_s^{(0)}+\hat{\mathcal{H}}'$. The non-interacting Hamiltonian corresponds to the KS Hamiltonian within BO
\begin{equation} \label{eq:H^0}
    \hat{\mathcal{H}}_s^{(0)} = \frac{\hat{p}^2}{2m}+ \hat{v}_{en}^{(0)}(\mathbf{r})
    + \hat{v}_{hxc}^{(0)}(\mathbf{r})\equiv\frac{\hat{p}^2}{2m}+\hat{v}_s^{(0)}(\mathbf{r})
    \;.
\end{equation}
Here $\hat{v}_{en}^{(0)}(\mathbf{r}) = \eval{\hat{v}_{en}(r,U)}_{U=0}$ is the electron-lattice Coulomb interaction evaluated at $\mathbf{R_0}$, and $\hat{v}_{hxc}^{(0)}(\mathbf{r})= \eval{\hat{v}_{hxc}(r,U)}_{U=0}$ corresponds to the standard HXC potential within BO, and
$\psi_{n\mathbf{k}}^{(0)}(\mathbf{r})$ and  $\epsilon_{n\mathbf{k}}^{(0)}(\mathbf{r})$ are, respectively, the KS eigenstates and eigenenergies that solve the unperturbed KS equation. 

We define the perturbation $\hat{\mathcal{H}}'$ as
\begin{equation}\label{eq:H'}
    \hat{\mathcal{H}}' = \hat{v}'_{s} +  
    \hat{v}_{geo}\;
\end{equation}
to second order in the phonon displacements, $U_{\mathbf{q}\lambda}$. Here, $\hat{v}'_{s}=\hat{v}_{s}^{(1)}+\hat{v}_{s}^{(2)}$ is the sum of the first- and second-order perturbation terms implemented in standard DFPT~\cite{gonze1995adiabatic,baroni2001phonons, giustino2017electron}, and  
$\hat{v}_{geo}=\hat{v}^{(1)}_{geo}+\hat{v}^{(2)}_{geo}$ is the \emph{non-adiabatic} correction. 
In the harmonic approximation, the nuclear wavefunctions are localized and have a zero-point amplitude of the order of $\mu^{1/4}$.  Hence, we can rationalize the orders of perturbation by recognizing~\cite{BO} that the characteristic nuclear displacement $U_{\mathbf{q}\lambda}$ is of order $\mu^{1/4}$.
The superscripts correspond to powers in the displacements. Crucially, we have two separate sources of $\mu$ dependence, on the one hand the $\mu$ dependence of the functional and on the other hand the $\mu$ dependence arising from the powers in the displacements $U$. The latter is already present in standard DFPT. 

We follow the first-order corrections previously derived in Ref.~\cite{RPG2019} for a single band case and extend them for the many-bands, many-mode case. Here, using the notations $1\equiv n_1,\mathbf{k_1}$ and $2\equiv n_2,\mathbf{k_2}$, $\hat{v}_{s}^{(1)}$ is defined as:
\begin{equation}\label{eq:v_KS1}
    \mel{\psi^{(0)}_{2}}{\hat{v}_{s}^{(1)}}{\psi^{(0)}_{1}}\equiv \sum_{\lambda} g_{n_2,n_1,\lambda}(\mathbf{k_1},\mathbf{q})\frac{U_{\mathbf{q}\lambda}}{L_{\mathbf{q}\lambda}} \;,
\end{equation}
where $\mathbf{k_2} = \mathbf{k_1}+\mathbf{q}$ as required by momentum conservation. $L_{\mathbf{q}\lambda}$ is the amplitude of the zero-point motion, defined as $L_{\mathbf{q}\lambda} = \sqrt{\frac{\hbar}{2M\omega_{\mathbf{q}\lambda}}}$, and $g_{n_2,n_1,\lambda}(\mathbf{k_1},\mathbf{q})$ is the first-order electron-phonon coupling term representing interactions between two electrons through a phonon. We define the first-order correction for $\hat{v}_{geo}^{(1)}$ using a Taylor expansion of the electronic wavefunctions. Following Eq.~\eqref{eq:v_geo_general},  this leads to
\begin{equation} \label{eq:v_geo1}
\begin{split}
    \mel{\psi_{2}^{(0)}}{\hat{v}_{geo}^{(1)}}{\psi_{1}^{(0)}} &=\sum_\lambda \hbar\omega_{\mathbf{q}\lambda} U_{\mathbf{q}\lambda} \braket{\psi_{2}}{\pdv{\psi^{(1)}_{1}}{U_{\mathbf{q}\lambda}}}\\
    &= \sum_\lambda g_{n_2,n_1,\lambda}(\mathbf{k_1},\mathbf{q})\frac{U_{\mathbf{q}\lambda}}{L_{\mathbf{q}\lambda}}\frac{\hbar\omega_{\mathbf{q}\lambda}}{\epsilon^{(0)}_{1}-\epsilon^{(0)}_{2}\pm\hbar\omega_{\mathbf{q}\lambda}}\;,
\end{split}
\end{equation}
where $\hbar\omega_{\mathbf{q}\lambda}$ is the phonon energy, accounting for both phonon emission and absorption channels. We note that these corrections are off-diagonal in the wavefunction basis, since they mix KS electron states upon phonon scattering. 
Since the first-order energy correction is defined as a diagonal term, and $\hat{v}_{s}^{(1)}$ and $\hat{v}_{geo}^{(1)}$ are off-diagonal by definition, this energy term vanishes, i.e. $\epsilon_{n\mathbf{k}}^{(1)}=0$.

The first-order correction to the wavefunctions follows the form
\begin{equation} \label{eq:1storder_wfn}
\begin{split}
    \ket{\psi_{1}^{(1)}}=& \sum_{2\neq 1} \frac{\mel{\psi^{(0)}_{2}}{\hat{v}_{s}^{(1)}+\hat{v}_{geo}^{(1)}}{\psi^{(0)}_{1}}}{\epsilon^{(0)}_{1}-\epsilon^{(0)}_{2}} \ket{\psi^{(0)}_{2}} \\
    =& \sum_\lambda \sum_{2\neq 1} \frac{g_{n_2,n_1,\lambda}(\mathbf{k_1},\mathbf{q})}{\epsilon^{(0)}_{1}-\epsilon^{(0)}_{2}\pm \hbar\omega_{\mathbf{q}\lambda}} \frac{U_{\mathbf{q}\lambda}}{L_{\mathbf{q}\lambda}} \;\ket{\psi^{(0)}_{2}}\;.
\end{split}
\end{equation}
Since we consider systems at zero temperature, only the phonon emission case appears hereafter.
These corrected wavefunctions can then be used to evaluate the changes in the density due to interaction with phonons. 
The electronic density, $n_U$, is defined as $n_U(\mathbf{r})=\sum_{n\mathbf{k}}f_{n\mathbf{k}U}\abs{\psi_{n\mathbf{k}U}}^2$, where $f_{n\mathbf{k}U}$ is the Fermi-Dirac occupation function for zero temperature. 
Hence, the first-order density correction is
\begin{equation} \label{eq:1st density}
\begin{split}
    n^{(1)}(\mathbf{r}) =\\ 
    2Re\sum_\lambda \sum_{1,2\neq 1}
    &\frac{g_{n_2,n_1,\lambda}(\mathbf{k_1},\mathbf{q})}{\epsilon_{1}^{(0)}-\epsilon_{2}^{(0)}-\hbar\omega_{\mathbf{q}\lambda}}
    \frac{U_{\mathbf{q}\lambda}}{L_{\mathbf{q}\lambda}} \psi_1^{(0)*}(\mathbf{r}) \psi_2^{(0)}(\mathbf{r})
\end{split}
\end{equation}

The general formula for the second-order energy correction is given by 
\begin{equation}\label{eq:en2_mel}
\begin{split}
    \epsilon^{(2)}_{1} =& \sum_{2\neq 1}\frac{\abs{\mel{\psi^{(0)}_{2}}{\hat{v}_{s}^{(1)}+\hat{v}_{geo}^{(1)}}{\psi^{(0)}_{1}}}^2}{\epsilon^{(0)}_{1}-\epsilon^{(0)}_{2}}\\
    &+\mel{\psi^{(0)}_{1}}{\hat{v}_{s}^{(2)}+\hat{v}_{geo}^{(2)}}{\psi^{(0)}_{1}}  \;.
\end{split}
\end{equation}
Here, the two-phonon-induced potentials $\hat{v}_s^{(2)}$ and $\hat{v}_{geo}^{(2)}$ appear. While the second-order correction to the energy eigenvalue $\epsilon_1^{(2)}$ only requires knowledge of the diagonal matrix elements of these potentials, since we ultimately seek second-order corrections to other observables such as the density, we proceed to evaluate all of the matrix elements of $\hat{v}_s^{(2)}$ and $\hat{v}_{geo}^{(2)}$.
In doing so, we need to take into account the gauge freedom in the choice of the electronic states, that is, the determinant of the occupied KS orbitals is invariant (up to a phase) to a unitary transformation among the occupied orbitals. 
Gonze (1995)~\cite{gonze1995adiabatic} has found a convenient gauge choice, called the ``parallel-transport'' gauge, for carrying out higher-order perturbative calculations.  Such a choice is specified through orthonormality constraints (as depicted in Eqs.~\eqref{eq-SI:orthonormal} and \eqref{eq-SI:paral_gauge}), which allow the calculations to be made with only the knowledge of the block off-diagonal components of the perturbations, i.e. the matrix elements between one occupied and one unoccupied state. As a second step, the results can be transformed back into the ``diagonal gauge'' - the familiar gauge choice in which the Lagrange multiplier matrix is diagonal with energy eigenvalues $\epsilon_1$ on the diagonal - producing corrections to the KS orbitals and eigenenergies which are consistent with the first-order corrections above. 

The matrix elements of the first-order perturbations $\hat{v}_{s}^{(1)}, \hat{v}_{geo}^{(1)}$ were addressed above. 
The matrix elements of second-order perturbations consist of the phonon-induced deviations of the KS potential and the non-adiabatic effect accounted for by $\hat{v}_{geo}$. 
Starting from the diagonal elements of $\hat{v}_s^{(2)}$ and $\hat{v}_{geo}^{(2)}$, required for the evaluation of Eq.~\eqref{eq:en2_mel}, we have
\begin{equation} \label{eq:V11-DW}
\begin{split}
    V^{(2)}_{1,1} \equiv \mel{\psi_{1}^{(0)}}{\hat{v}_{s}^{(2)}}{\psi_{1}^{(0)}} = 
    \sum_{\mathbf{q}\lambda} g^{(2)}_{n_1,n_1;\lambda,\lambda}(\mathbf{k}_1,\mathbf{q},\bar{\mathbf{q}}) \frac{\abs{U_{\mathbf{q}\lambda}}^2}{L_{\mathbf{q}\lambda}^2}\;,
\end{split}
\end{equation}
$V^{(2)}_{1,1}$ is the potential perturbation that accounts for the two-phonon process originating from the Debye-Waller (DW) interaction $g^{(2)}_{n_1,n_1;\lambda,\lambda}(\mathbf{k}_1,\mathbf{q},\bar{\mathbf{q}})$, which can be obtained in adiabatic DFPT calculations~\cite{PONCE2014341,giustino2017electron,cardona2005isotope,allen1976theory,allen1981theory,giustino2010electron,Park2020}. 
The expression for the $\hat{v}^{(2)}_{geo}$ matrix element, which follows from Eq.~\eqref{eq:v_geo_general} in the parallel-transport gauge (derivation elaborated in appendix~\ref{app:A}), amounts to
\begin{equation}  \label{eq:W11}
    \begin{split}
        &W^{(2)}_{1,1} \equiv \mel{\psi_{1}^{(0)}}{\hat{v}_{geo}^{(2)}}{\psi_{1}^{(0)}}\\ &= \sum_{2}^{unocc} \sum_{\lambda\lambda'}\frac{g^*_{n_2,n_1,\lambda'}(\mathbf{k_1},\mathbf{q})g_{n_2,n_1,\lambda}(\mathbf{k_1},\mathbf{q})\hbar\omega_{\mathbf{q}\lambda}}{(\epsilon_1^{(0)}-\epsilon_2^{(0)}-\hbar\omega_{\mathbf{\bar{q}}\lambda'})(\epsilon_1^{(0)}-\epsilon_2^{(0)}-\hbar\omega_{\mathbf{q}\lambda})}\\
        &\cross\bigg(\delta_{\lambda\lambda'} - \frac{U_{\mathbf{\bar{q}}\lambda'}}{L_{\mathbf{\bar{q}}\lambda'}}\frac{U_{\mathbf{q}\lambda}}{L_{\mathbf{q}\lambda}} \bigg)\\
        &-\sum_{2}^{unocc}\sum_{\lambda\lambda'}\frac{g^*_{n_2,n_1,\lambda'}(\mathbf{k_1},\mathbf{q})g_{n_2,n_1,\lambda}(\mathbf{k_1},\mathbf{q})\hbar\omega_{\mathbf{q}\lambda}}{(\epsilon_1^{(0)}-\epsilon_2^{(0)}-\hbar\omega_{\mathbf{\bar{q}}\lambda'})(\epsilon_1^{(0)}-\epsilon_2^{(0)}-\hbar\omega_{\mathbf{q}\lambda})}\frac{U_{\mathbf{\bar{q}}\lambda'}}{L_{\mathbf{\bar{q}}\lambda'}}\frac{U_{\mathbf{q}\lambda}}{L_{\mathbf{q}\lambda}}{,}
    \end{split}
\end{equation}
where $\mathbf{q}=\mathbf{k}_2-\mathbf{k_1}$ and $\overline{\mathbf{q}}=-\mathbf{q}$.

When the quantity in Eq.~\eqref{eq:W11} is multiplied by $\abs{\chi(U_{\mathbf{q}\lambda})}^2$ and averaged over the phonon amplitudes, the first line 
cancels out. 
Using the definition for the matrix elements in Eqs.~\eqref{eq:v_KS1}, \eqref{eq:v_geo1}, \eqref{eq:V11-DW} and \eqref{eq:W11} and transforming the second-order eigenenergy correction to the diagonal gauge, as shown in appendix \ref{app:B}, the energy correction results in  
\begin{equation} \label{eq:eps2_U}
    \begin{split}
       &\epsilon_{1}^{(2)} =\\ 
       &\sum_{2\neq 1} \sum_{\lambda\lambda'}  \bigg[\frac{\abs{g_{n_2,n_1,\lambda}(\mathbf{k_1},\mathbf{q})}^2}{\epsilon^{(0)}_{1}-\epsilon^{(0)}_{2}-\hbar\omega_{\mathbf{q}\lambda}} \frac{\abs{U_{\mathbf{q}\lambda}}^2}{L^2_{\mathbf{q}\lambda}}\\
       &+ \frac{g^*_{n_2,n_1,\lambda'}(\mathbf{k_1},\mathbf{q})g_{n_2,n_1,\lambda}(\mathbf{k_1},\mathbf{q}) \hbar\omega_{\mathbf{q}\lambda}}{(\epsilon^{(0)}_{1}-\epsilon^{(0)}_{2}-\hbar\omega_{\bar{\mathbf{q}}\lambda'})(\epsilon^{(0)}_{1}-\epsilon^{(0)}_{2}-\hbar\omega_{\mathbf{q}\lambda})}\\
       &\cross \left(\delta_{\lambda\lambda'}- \frac{{U_{\bar{\mathbf{q}}\lambda'}}}{L_{\bar{\mathbf{q}}\lambda'}}\frac{{U_{\mathbf{q}\lambda}}}{L_{\mathbf{q}\lambda}}\right)\bigg]+\sum_{\mathbf{q}\lambda} g^{(2)}_{n_1,n_1;\lambda,\lambda}(\mathbf{k}_1,\mathbf{q},\bar{\mathbf{q}}) \frac{\abs{U_{\mathbf{q}\lambda}}^2}{L_{\mathbf{q}\lambda}^2}\;. 
    \end{split}
\end{equation}

Note that conventionally the electronic band structure is evaluated for fixed nuclear positions (within the BO approximation), however in the current framework it explicitly depends on the displacement.
In the context of EF any purely electronic observable naturally comes out as a $\abs{\chi(U_{\mathbf{q}\lambda})}^2$ average. 
To make contact with the standard Fan-Migdal (FM) and DW result, we average over $U_{\mathbf{q}\lambda}$ using $\abs{\chi(U_{\mathbf{q}\lambda})}^2$ as a weighting function. As before, the middle term of Eq.~\eqref{eq:eps2_U} cancels out.
The average energy renormalization is
\begin{equation} \label{eq:eps2_avarage}
    \begin{split}
       \bar{\epsilon}_{1}^{(2)} = \sum_{2\neq 1} \sum_{\lambda}  \frac{\abs{g_{n_2,n_1,\lambda}(\mathbf{k_1},\mathbf{q})}^2}{\epsilon^{(0)}_{1}-\epsilon^{(0)}_{2}-\hbar\omega_{\mathbf{q}\lambda}} 
       + \sum_{\mathbf{q}\lambda}
      g^{(2)}_{n_1,n_1;\lambda,\lambda}(\mathbf{k}_1,\mathbf{q},\bar{\mathbf{q}})\;.
    \end{split}
\end{equation}
The first term consists of the electron-one phonon interaction and the associated non-adiabatic corrections. This reproduces the energy renormalization due to the well-known FM self-energy~\cite{migdal1958interaction,giustino2017electron,RPG2019}, dressed with non-adiabatic interactions. The second term consists of the electron-two phonon interaction.
In Eqs.~\eqref{eq:eps2_U} and \eqref{eq:eps2_avarage} there are second-order terms with non-adiabatic corrections entering through the dependence on phonon frequencies, as well as additional dependence on $\abs{U_{\mathbf{q}\lambda}}^2/L_{\mathbf{q}\lambda}^2$. On the other hand, at this order there are no non-adiabatic corrections to the last term of Eq.~\eqref{eq:eps2_avarage} originating from the DW interaction. However, there will be corrections to the off-diagonal DW terms, as we discuss next.

We continue to the evaluation of the off-diagonal elements of $\hat{v}_{s}^{(2)}$ and $\hat{v}_{geo}^{(2)}$, which are required for the second-order wavefunction corrections. 
The expression for $\hat{v}_{s}^{(2)}$ elements follows

\begin{equation}\label{eq:V2}
    \begin{split}
        V^{(2)}_{2,1} &\equiv \mel{\psi_{2}^{(0)}}{\hat{v}_{s}^{(2)}}{\psi_{1}^{(0)}}\\ 
        &=\sum_{\mathbf{q}\lambda\lambda'} g^{(2)}_{n_2,n_1,\lambda,\lambda'}(\mathbf{k}_1,\mathbf{q},\mathbf{q}') \frac{U_{\mathbf{q}\lambda}}{L_{\mathbf{q}\lambda}} \frac{U_{\mathbf{q}'\lambda'}}{L_{\mathbf{q}'\lambda'}}{,}
    \end{split}
\end{equation}

where $\mathbf{q}'=\mathbf{k}_2-\mathbf{k_1}-\mathbf{q}$, $1\in$ occupied states subspace and $2\in$ unoccupied states subspace.
The expression for the $\hat{v}^{(2)}_{geo}$ off-diagonal matrix element in the parallel-transport gauge, allows for the construction of a differential equation for $W^{(2)}_{2,1}$ (derivation elaborated in appendix~\ref{app:A}),
\begin{equation}\label{eq:W2_mel_ODE}
\begin{split}
    &W^{(2)}_{2,1} \equiv \mel{\psi_{2}^{(0)}}{\hat{v}_{geo}^{(2)}}{\psi_{1}^{(0)}}=\sum_{\mathbf{q}\lambda}\hbar\omega_{\mathbf{q}\lambda}U_{\mathbf{q}\lambda}\braket{\psi_{2}^{(0)}}{\pdv{\psi_{1}^{(2)}}{U_{\mathbf{q}\lambda}}}\\
    & -\delta_{\mathbf{k_2},\mathbf{k_1}}\sum_{\mathbf{q}\lambda}\hbar\omega_{\mathbf{q}\lambda}L_{\mathbf{q}\lambda}L_{\mathbf{\bar{q}}\lambda}\braket{\psi_{2}^{(0)}}{\pdv{\psi_{1}^{(2)}}{U_{\mathbf{\bar{q}}\lambda}}{U_{\mathbf{q}\lambda}}}\;.
\end{split}
\end{equation}

With the evaluation of $\braket{\psi_{2}^{(0)}}{\psi_{1}^{(2)}}$ from the second-order Sternheimer equation with the additional non-adaiabatic potential $\hat{v}_{geo}^{(2)}$ and using the notation of $V^{(2)}_{2,1}, W^{(2)}_{2,1}$ for the second-order matrix elements, the differential equation becomes
\begin{equation}\label{eq:W2_mel_ODE_WV}
\begin{split}
    &W^{(2)}_{2,1} =\\
    &\sum_{\mathbf{q}\lambda}\hbar\omega_{\mathbf{q}\lambda}U_{\mathbf{q}\lambda}\frac{1}{\epsilon_{1}^{(0)}-\epsilon_{2}^{(0)}} \pdv{(V^{(2)}_{2,1}+W^{(2)}_{2,1})}{U_{\mathbf{q}\lambda}}\\
    &-\delta_{\mathbf{k_2}\mathbf{k_1}}\sum_{\mathbf{q}\lambda}\hbar\omega_{\mathbf{q}\lambda}L_{\mathbf{q}\lambda}L_{\mathbf{\bar{q}}\lambda} \frac{1}{\epsilon_{1}^{(0)}-\epsilon_{2}^{(0)}}\pdv{(V^{(2)}_{2,1}+W^{(2)}_{2,1})}{U_{\mathbf{\bar{q}}\lambda}}{U_{\mathbf{q}\lambda}} \;.
\end{split}
\end{equation}

The derivation from here to the explicit form of the differential equation is given in appendix~\ref{app:A}, leading to 
\begin{equation}\label{eq:W2_ODE}
\begin{split}
    &W^{(2)}_{n_2\mathbf{k_2},n_1\mathbf{k_1}}=\\  &\sum_{\mathbf{q}\lambda\lambda'}\frac{\hbar\omega_{\mathbf{q}\lambda}}{\epsilon_{n_1\mathbf{k_1}}^{(0)}-\epsilon_{n_2\mathbf{k_2}}^{(0)}}\left(
    \frac{U_{\mathbf{q}\lambda}}{L_{\mathbf{q}\lambda}}\frac{U_{\mathbf{q}'\lambda'}}{L_{\mathbf{q}'\lambda'}} - \delta_{\mathbf{k}_2\mathbf{k}_1}\delta_{\lambda\lambda'} \right)\\ &\cross
    \big[g^{(2)}_{n_2,n_1,\lambda,\lambda'}(\mathbf{k_1},\mathbf{q},\mathbf{q}') +g^{(2)}_{n_2,n_1,\lambda',\lambda}(\mathbf{k_1},\mathbf{q}',\mathbf{q}) \big]
    \\
    &+ \sum_{\mathbf{q}\lambda}\frac{\hbar\omega_{\mathbf{q}\lambda}}{\epsilon_{n_1\mathbf{k_1}}^{(0)}-\epsilon_{n_2\mathbf{k_2}}^{(0)}} U_{\mathbf{q}\lambda}\frac{\partial W^{(2)}_{n_2\mathbf{k_2},n_1\mathbf{k_1}}}{\partial U_{\mathbf{q}\lambda}}\\
    &-\delta_{\mathbf{k_2}\mathbf{k_1}}\sum_{\mathbf{q}\lambda}\frac{\hbar\omega_{\mathbf{q}\lambda}}{\epsilon_{n_1\mathbf{k_1}}^{(0)}-\epsilon_{n_2\mathbf{k_2}}^{(0)}}L_{\mathbf{q}\lambda}L_{\mathbf{\bar{q}}\lambda}\frac{\partial^2 W^{(2)}_{n_2\mathbf{k_2},n_1\mathbf{k_1}}}{\partial U_{\mathbf{\bar{q}}\lambda}\partial U_{\mathbf{q}\lambda}} \;,
\end{split}
\end{equation}
where $\mathbf{q}'=\mathbf{k}_2-\mathbf{k}_1-\mathbf{q}$.
The general solution of $\hat{v}_{geo}^{(2)}$ can include non-analytical terms or polynomials that reflect non-analytic behavior and meaningful physical constraints. Here we introduce an analytical expression and reduce our solution to a particular case.  
A particular solution to the differential equation in terms of standard \textit{ab initio} calculated quantities is
\begin{equation}\label{eq:W2_sol}
\begin{split}
    &W^{(2)}_{n_2\mathbf{k_2},n_1\mathbf{k_1}} =\\ &\sum_{\mathbf{q}\lambda\lambda'} \frac{1}{2}
    \left(g^{(2)}_{n_2,n_1,\lambda,\lambda'}(\mathbf{k_1},\mathbf{q},\mathbf{q}') +g^{(2)}_{n_2,n_1,\lambda',\lambda}(\mathbf{k_1},\mathbf{q}',\mathbf{q}) \right)\\
    &\cross \frac{\hbar\omega_{\mathbf{q}\lambda}+\hbar\omega_{\mathbf{q}',\lambda'}}{\epsilon_{n_1\mathbf{k_1}}^{(0)}-\epsilon_{n_2\mathbf{k_2}}^{(0)}-\hbar \omega_{\mathbf{q}\lambda}-\hbar\omega_{\mathbf{q}'\lambda'}}\\
    &\cross\left(\frac{U_{\mathbf{q}\lambda}}{L_{\mathbf{q}\lambda}}\frac{U_{\mathbf{q}'\lambda'}}{L_{\mathbf{q}'\lambda'}} - \delta_{\mathbf{k}_1\mathbf{k}_2} \delta_{\lambda\lambda'}\right)\;.
\end{split}
\end{equation}

Finally, we demonstrate the orbital changes in this level of approximation.
The second-order wavefunction correction in the parallel-transport gauge reads 
\begin{equation}\label{SI-eq:2_mel}
\begin{split}
    \ket{\psi_{1,||}^{(2)}} =& 
    -\frac{1}{2} \sum_{3\neq 1}^{occ}\sum_{2 \neq 1}^{unocc} 
    \frac{\mel{\psi^{(0)}_{3}}{(\hat{v}_{s}^{(1)}+\hat{v}_{geo}^{(1)})}{\psi^{(0)}_{2}} }{(\epsilon_{3}^{(0)}-\epsilon_{2}^{(0)})} \\
    & \cross \frac{\mel{\psi^{(0)}_{2}}{(\hat{v}_{s}^{(1)}+\hat{v}_{geo}^{(1)})}{\psi^{(0)}_{1}}}{(\epsilon_{1}^{(0)}-\epsilon_{2}^{(0)})}\ket{\psi^{(0)}_{3}}\\
    &-\sum_{3\neq 1}^{unocc} \frac{\mel{\psi^{(0)}_{3}}{\hat{v}_{s}^{(2)}+\hat{v}_{geo}^{(2)}}{\psi^{(0)}_{1}}}{(\epsilon_{3}^{(0)}-\epsilon_{1}^{(0)})} \ket{\psi^{(0)}_{3}}\\
    &-\frac{1}{2}\sum_{2\neq 1}^{unocc} \frac{\abs{\mel{\psi^{(0)}_{2}}{(\hat{v}_{s}^{(1)}+\hat{v}_{geo}^{(1)})}{\psi^{(0)}_{1}}}^2}{(\epsilon_{2}^{(0)}-\epsilon_{1}^{(0)})^2} \ket{\psi^{(0)}_{1}}\;.
\end{split}
\end{equation}
The correction to the single-particle KS states in the diagonal gauge is evaluated as
\begin{equation}\label{eq:psi||2_trans}
    \ket{\psi^{(2)}_{1,d}} = \ket{\psi^{(2)}_{1,||}}+\sum_3^{occ} \mathcal{U}_{13}^{(1)*}\ket{\psi^{(1)}_{3,||}}+\sum_3^{occ} \mathcal{U}_{13}^{(2)*}\ket{\psi^{(0)}_{3}},
\end{equation}
following Gonze~\cite{gonze1995adiabatic} and the derivation in appendix~\ref{app:C}.
This shows that second-order wavefunction corrections account for modifications in the electronic KS states due to the two-phonon interaction. This marks a step toward a self-consistent description of the solution, including mutual NA phonon-induced interactions between multiple electronic states.

\section{Discussion}

In the EF-DFPT method presented here we achieve explicit expressions correcting the KS wavefunctions and band structure to second order in the displacement due to non-adiabatic effects. The current method is rigorously derived from the full Hamiltonian of electrons and nuclei, in contrast to the commonly used model or Fr\"ohlich Hamiltonian starting point. Thus in principle it accounts for all non-adiabatic effects. 

There is further sensitivity in the choice of the perturbative treatment applied to the Hamiltonian. Here we expand all quantities of the GKS Hamiltonian and the nuclear Hamiltonian systematically in powers of the electron over nuclei mass ratio. In principle, all orders of the mass ratio in the EF potential can be accounted for. The EF-KS Hamiltonian explicitly depends on the nuclear wavefunction through the introduction of $\hat{v}_{geo}$, which is a higher order term in the mass ratio compared to the standard KS potential. The corrected wavefunctions are conditional electronic states. For localized nuclear wavefunctions, the characteristic zero-point amplitude is of the order $\mu^{1/4}$, which implies that the nuclear displacements should be assigned to the same order in standard DFPT. This is a way for the mass dependence to enter in DFPT because the BO potential energy surface and the KS potential have no explicit dependence on the mass ratio. Therefore, DFPT introduces some non-adiabatic effects whereas EF has systematically all orders of mass ratio.

We emphasize that the corrections to the electronic wavefunctions derived above are in fact corrections to the conditional quantities, i.e. the KS orbitals for a given nuclear configuration. Consequently, the density evaluated from these corrected wavefunctions is also the conditional one. An important feature of the presented approach is the ability to quantify the non-adiabatic contribution in the corrections explicitly. This is reflected in the conditional wavefunctions and density as well as in any observable evaluated as a function of these quantities,
such as the transition dipole moment in optical absorption, the total energy, electron and spin polarization, the dielectric screening function, and the electronic self-energy, extending beyond their evaluation at the DFT level, as well as in the update of the nuclear degrees of freedom.
  
It is possible to relate the band structure corrections to self-energy corrections in Green's function based methods, and associate the energy corrections derived here with a non-adiabatic treatment of the Allen-Heine-Cardona (AHC) theory~\cite{allen1976theory}. The relation of the wavefunctions corrections to Green's function based approaches is less straightforward. Although both the KS wavefunctions and Green's functions can be used for the evaluation of the electronic density, it is crucial to stress that in the current context the density evaluated is the conditional one. Hence a potential mapping between the different approaches requires introduction of nuclear configuration dependence into the electronic Green's function, or potentially casting the EF formalism into a many-body Green's function theory~\cite{harkonen2022exact}.

We further note that in the current derivation, temperature effects are not taken into account. These can be added through the electron Fermi-Dirac occupation function, $f_{n\mathbf{k}}$, and the phonon Bose-Einstein occupation function, $N_{\mathbf{q}\lambda}$, e.g. in the density correction, Eq.~\eqref{eq:1st density}~\cite{caruso2021nonequilibrium}. The theoretical framework presented here is designed in principle for ground-state properties, and the introduction of thermal excitation as reflected in finite temperature should be addressed carefully. 
The extension of this approach to include thermal excitations and fluctuations in the occupation in a time-dependent formalism may provide an insight into non-adiabatic dynamical properties in the system.

\section{Conclusions}

We have developed an \textit{ab initio} EF-based framework for the evaluation of the non-adiabatic signatures of electron-lattice interaction in real materials. 
It allows for the explicit identification of these signatures in the electronic conditional band structure, wavefunction, density, and properties derived from them, framing them in terms of DFPT calculated quantities.
Eventually, the perturbative corrections may be used iteratively to address the self-consistent solution of the nuclear and the electronic set of equations. Thus, this work takes a first step towards the full EF treatment of the electron-lattice many-body problem in materials.

\begin{acknowledgments}
\vspace{-10pt}
G.C. acknowledges an Institute for Environmental Sustainability (IES) Fellowship. 
E.K.U.G. acknowledges support as Mercator Fellow within SFB 1242 at the University Duisburg-Essen.
This project has received funding from the European Research Council (ERC) under the European Union's Horizon 2020 research and innovation programme, grant agreements No. ERC-2017-AdG-788890 and No.ERC-2022-StG-101041159, and from the Israel Science Foundation, grant agreement No. 1208/19.
\end{acknowledgments}

\appendix
\section{}\label{app:A}

Below we derive the $\hat{v}_{geo}^{(2)}$ matrix elements, using the notations $1\equiv n_1,\mathbf{k_1}$ and $2\equiv n_2,\mathbf{k_2}$.
The expression for $\hat{v}_{geo}$ is given in Eq.~\eqref{eq:v_geo_general}
with the orbital-dependent functional~\cite{RPG2019},
\begin{equation}
    \varepsilon_{geo} = \frac{\hbar^2}{2M}\sum_{\mathbf{q}\lambda}\sum_1^{occ} \mel{\frac{\partial \psi_1}{\partial U_{\mathbf{q}\lambda}}}{(1-P_v)}{\frac{\partial \psi_1}{\partial U_{\mathbf{q}\lambda}}}{,}
\end{equation}
where $(1-P_v)=P_c=\sum_3^{unocc}|\psi_3\rangle\langle\psi_3|$ is the projector onto the unoccupied subspace.
By taking the functional derivative of $\varepsilon_{geo}$ term in the total energy, the $\hat{v}_{geo}$ matrix element given in Eq.~\eqref{eq:v_geo_general} becomes 
\begin{equation}\label{SI-eq:v_geo_full}
\begin{split}
    &\mel{\psi_2}{\hat{v}_{geo}}{\psi_1} =\\
    &-\frac{\hbar^2}{2M}\sum_{\mathbf{q}\lambda}\sum_3^{occ} \braket{\psi_2}{\frac{\partial \psi_3}{\partial U_{\mathbf{q}\lambda}}}\braket{\frac{\partial \psi_3}{\partial U_{\mathbf{q}\lambda}}}{\psi_1}   \;\;\;\;\;\;\;\;\;(\romannumeral1)\\
    &-\frac{\hbar^2}{2M}\sum_{\mathbf{q}\lambda}\sum_1^{occ} \frac{\partial \ln{\abs{\chi}^2}}{\partial U^*_{\mathbf{q}\lambda}}\mel{\psi_2}{(1-P_v)}{\frac{\partial \psi_1}{\partial U_{\mathbf{q}\lambda }}}  \;(\romannumeral2)\\
    &-\frac{\hbar^2}{2M}\sum_{\mathbf{q}\lambda}\sum_1^{occ}\mel{\psi_2}{(1-P_v)}{\frac{\partial^2 \psi_1}{\partial U^*_{\mathbf{q}\lambda} \partial U_{\mathbf{q}\lambda}}}  \;\;\;\;\;(\romannumeral3)\\
    &+\frac{\hbar^2}{2M}\sum_{\mathbf{q}\lambda}\braket{\frac{\partial \psi_2}{\partial U_{\mathbf{q}\lambda}}}{\frac{\partial \psi_1}{\partial U_{\mathbf{q}\lambda}}} \;\;\;\;\;\;\;\;\;\;\;\;\;\;\;\;\;\;\;\;\;\;\;\;\;\;\;\;\;\,(\romannumeral4)\\
    &+\frac{\hbar^2}{2M}\sum_{\mathbf{q}\lambda}\sum_3^{occ} \braket{\psi_2}{\frac{\partial \psi_3}{\partial U_{\mathbf{q}\lambda}^*}}\braket{\psi_3}{\frac{\partial \psi_1}{\partial U_{\mathbf{q}\lambda}}}  \;\;\;\;\;\;\;\;\;(\romannumeral5)
\end{split}
\end{equation}

The second-order contribution to the matrix element of $\hat{v}_{geo}$  is given by the sum of three terms. The first is the matrix element of the second-order component of $\hat{v}_{geo}^{(2)}$ between the unperturbed wavefunctions. The second and third are the matrix elements of the first-order component $\hat{v}_{geo}^{(1)}$ between an unperturbed state and the first-order corrected state. 

\begin{equation}\label{eq:psi_vgeo_psi_2}
\begin{split}
    &\bra{\psi_2}\hat{v}_{geo}\ket{\psi_1}^{(2)} =\\ &\bra{\psi_2^{(0)}}\hat{v}_{geo}^{(2)}\ket{\psi_1^{(0)}}+ \bra{\psi_2^{(0)}}\hat{v}_{geo}^{(1)}\ket{\psi_1^{(1)}} +\bra{\psi_2^{(1)}}\hat{v}_{geo}^{(1)}\ket{\psi_1^{(0)}}\;.
\end{split}
\end{equation}

The perturbed KS states are subjected to orthonormalization conditions, 
\begin{equation}\label{eq-SI:orthonormal}
\begin{split}
    \braket{\psi_{2}^{(0)}}{\psi_{1}^{(i)}}&+\braket{\psi_{2}^{(i)}}{\psi_{1}^{(0)}} =\\ 
    &\begin{cases}
        -\sum^{i-1}_{j=1}\braket{\psi_{2}^{(j)}}{\psi_{1}^{(i-j)}} & i>1\\
        0 & i=1\;,
    \end{cases}
\end{split}
\end{equation}
where $i$ is the order of the perturbation, $1,2$ are either both occupied or both unoccupied states.
The additional constraints defining the parallel-transport gauge~\cite{gonze1995adiabatic} are
\begin{equation}\label{eq-SI:paral_gauge}
    \braket{\psi_{2}^{(0)}}{\psi_{1}^{(i)}}-\braket{\psi_{2}^{(i)}}{\psi_{1}^{(0)}}=0\;.
\end{equation}

\subsection{Diagonal elements of $\mathbf{\hat{v}_{geo}^{(2)}}$}
Using this gauge choice and expanding the terms in Eq.~\eqref{SI-eq:v_geo_full} for $\mel{\psi_1}{\hat{v}_{geo}}{\psi_1}$, up to second order in the displacements $U_{\mathbf{q}\lambda}$, it is possible to show that terms $(\romannumeral1)$ and $(\romannumeral5)$, and terms $(\romannumeral2)$, $(\romannumeral3)$, do not contribute at the second order. Thus, this choice of gauge allows a convenient representation of $\hat{v}_{geo}^{(2)}$. 
To show this explicitly, we first focus on  (\romannumeral1) which can be written as
\begin{equation}
    (\romannumeral1)= -\frac{\hbar^2}{2M}\sum_{\mathbf{q}\lambda}\sum_3^{occ} \braket{\frac{\partial \psi_1}{\partial U_{\mathbf{q}\lambda}}}{\psi_3}\braket{\psi_3}{\frac{\partial \psi_1}{\partial U_{\mathbf{q}\lambda}}}\;.
\end{equation}
Expanding the last factor we obtain
\begin{equation}
\begin{split}
    \braket{\psi_3}{\frac{\partial \psi_1}{\partial U_{\mathbf{q}\lambda}}}
    &= \braket{\psi_3^{(0)}}{\frac{\partial \psi_1^{(0)}}{\partial U_{\mathbf{q}\lambda}}} +\braket{\psi_3^{(0)}}{\frac{\partial \psi_1^{(1)}}{\partial U_{\mathbf{q}\lambda}}} + \order{U^2}\;.
\end{split}
\end{equation}
In this term $1,3 \in$ occupied subspace. The derivative of the unperturbed wavefunction with respect to the displacement is zero, hence the first term vanishes. Due to the orthogonality of the basis and the parallel-transport gauge, Eqs.~\eqref{eq-SI:orthonormal} and \eqref{eq-SI:paral_gauge}, it follows that the second term vanishes as well and this factor results in terms of $\order{U^2}$. Therefore, since the first factor $\braket{\frac{\partial \psi_2}{\partial U_{\mathbf{q}\lambda}}}{\psi_3}$ is also $\order{U^2}$, the (\romannumeral1) term is of order $\order{U^4}$. Similarly, (\romannumeral5) is also of order $\order{U^4}$ and therefore they do not contribute at second order.

The diagonal elements in terms $(\romannumeral2)$ and $(\romannumeral3)$ consist of $\bra{\psi_1}(1-P_v)$
for $1\in$ occupied subspace.
The projector, as well as $\bra{\psi_1}$, contain all orders of the perturbation. Therefore  $(\romannumeral2)$ and $(\romannumeral3)$ vanish due to orthonormality. 

The only contribution to the second-order diagonal elements comes from $(\romannumeral4)$. After expanding the wavefunctions in 
$(\romannumeral4)$ and considering the gauge choice, the contributing elements are constructed from the first-order perturbed wavefunction $\psi_1^{(1)}$ (expressed in Eq.~\eqref{eq:1storder_wfn}). Thus,
\begin{equation}
\begin{split}
        &\mel{\psi_1}{\hat{v}_{geo}}{\psi_1}^{(2)} =   \frac{\hbar^2}{2M} \sum_{\mathbf{q}\lambda}\braket{\frac{\partial \psi^{(1)}_1}{\partial U_{\mathbf{q}\lambda}}}{\frac{\partial \psi^{(1)}_1}{\partial U_{\mathbf{q}\lambda}}}\\
        &= \sum_{\lambda}\sum_2^{unocc}\frac{g^*_{n_2,n_1,\lambda}(\mathbf{k_1},\mathbf{q})g_{n_2,n_1,\lambda}(\mathbf{k_1},\mathbf{q})\hbar\omega_{\mathbf{q}\lambda}}{(\epsilon_1^{(0)}-\epsilon_2^{(0)}-\hbar\omega_{\mathbf{q}\lambda})^2}\;.
\end{split}
\end{equation}
Note that the degree of freedom of $\mathbf{q}$ is accounted for by the sum over $2\equiv n_2,\mathbf{k_2}=n_2,\mathbf{k_1}+\mathbf{q}$.
From here we can derive the diagonal matrix element $W^{(2)}_{11} \equiv \bra{\psi_1^{(0)}}\hat{v}_{geo}^{(2)}\ket{\psi_1^{(0)}}$ as
\begin{equation}
\begin{split}
    &W^{(2)}_{11} =\\ 
    &\bra{\psi_1}\hat{v}_{geo}\ket{\psi_1}^{(2)}-\bra{\psi_1^{(0)}}\hat{v}_{geo}^{(1)}\ket{\psi_1^{(1)}}-\bra{\psi_1^{(1)}}\hat{v}_{geo}^{(1)}\ket{\psi_1^{(0)}}\\
    &= \sum_{\lambda}\sum_2^{unocc}\frac{g^*_{n_2,n_1,\lambda}(\mathbf{k_1},\mathbf{q})g_{n_2,n_1,\lambda}(\mathbf{k_1},\mathbf{q})\hbar\omega_{\mathbf{q}\lambda}}{(\epsilon_1^{(0)}-\epsilon_2^{(0)}-\hbar\omega_{\mathbf{q}\lambda})^2}\\
    &- \sum_{\lambda\lambda'}\sum_2^{unocc}\frac{g^*_{n_2,n_1,\lambda'}(\mathbf{k_1},\mathbf{q})g_{n_2,n_1,\lambda}(\mathbf{k_1},\mathbf{q})(\hbar\omega_{\mathbf{\bar{q}}\lambda'}+\hbar\omega_{\mathbf{q}\lambda})}{(\epsilon_1^{(0)}-\epsilon_2^{(0)}-\hbar\omega_{\mathbf{\bar{q}}\lambda'})(\epsilon_1^{(0)}-\epsilon_2^{(0)}-\hbar\omega_{\mathbf{q}\lambda})}\\&\cross\frac{U_{\mathbf{\bar{q}}\lambda'}}{L_{\mathbf{\bar{q}}\lambda'}}\frac{U_{\mathbf{q}\lambda}}{L_{\mathbf{q}\lambda}}\;.
\end{split}
\end{equation}
Since the factor
\begin{equation}
    \frac{g^*_{n_2,n_1,\lambda'}(\mathbf{k_1},\mathbf{q})g_{n_2,n_1,\lambda}(\mathbf{k_1},\mathbf{q})}{(\epsilon_1^{(0)}-\epsilon_2^{(0)}-\hbar\omega_{\mathbf{\bar{q}}\lambda'})(\epsilon_1^{(0)}-\epsilon_2^{(0)}-\hbar\omega_{\mathbf{q}\lambda})}\frac{U_{\Bar{q}\lambda'}}{L_{\mathbf{\bar{q}}\lambda'}}\frac{U_{\mathbf{q}\lambda}}{L_{\mathbf{q}\lambda}}
\end{equation}
is symmetric with respect to interchanging $\mathbf{q}\lambda$ with $\mathbf{\bar{q}}\lambda'$ we finally get the diagonal elements of $\hat{v}_{geo}^{(2)}$ as in Eq.~\eqref{eq:W11}.

\subsection{Off-diagonal elements of $\mathbf{\hat{v}_{geo}^{(2)}}$}
To work out the off-diagonal terms of $\hat{v}_{geo}^{(2)}$ we turn back to Eq.~\eqref{SI-eq:v_geo_full} in its off-diagonal form and derive it to second order. From the same considerations elaborated above terms $(\romannumeral1)$ and $(\romannumeral5)$ do not contribute at this order.
Term $(\romannumeral4)$, which contributes to the diagonal part of $\hat{v}_{geo}^{(2)}$, does not contribute to the off-diagonal elements due to the gauge choice. 

The remaining terms are (\romannumeral2) and (\romannumeral3). 
As before, these terms consist of $\bra{\psi_2}(1-P_v)$, where here $2\in$ unoccupied subspace, in contrast to the diagonal case. Therefore, these terms do not vanish. 
In the current case $\bra{\psi_2}(1-P_v)$ amounts to the unperturbed wavefunction $\bra{\psi_2^{(0)}}$. 
The second-order contribution of $(\romannumeral2)$ results in
\begin{equation}
\begin{split}
    &-\frac{\hbar^2}{2M}\sum_{\mathbf{q}\lambda} \frac{\partial \ln{\abs{\chi}^2}}{\partial U^*_{\mathbf{q}\lambda}}\braket{\psi_2^{(0)}}{\pdv{\psi_1}{U_{\mathbf{q}\lambda }}}\\
    &=\frac{\hbar^2}{2M} \sum_{\mathbf{q}\lambda} \frac{U_{\mathbf{q}\lambda}}{L_{\mathbf{q}\lambda}^2}
    \pdv{\braket{\psi_2^{(0)}}{\psi_1^{(2)}}}{U_{\mathbf{q}\lambda}}\\
    & =  \sum_{\mathbf{q}\lambda} \hbar\omega_{\mathbf{q}\lambda} U_{\mathbf{q}\lambda} \pdv{\braket{\psi_2^{(0)}}{\psi_1^{(2)}}}{U_{\mathbf{q}\lambda}} \;.
\end{split}
\end{equation}
Similarly, the second-order contribution from $(\romannumeral3)$, follows
\begin{equation}
    -\frac{\hbar^2}{2M} \sum_{\mathbf{q}\lambda} \sum_{n_2}^{unocc} \delta_{\mathbf{k_2},\mathbf{k_1}} \pdv{\braket{\psi_2^{(0)}}{\psi_1^{(2)}}}{U^*_{\mathbf{q}\lambda}}{U_{\mathbf{q}\lambda}}  \;.
\end{equation}

These contributions add up to form the off-diagonal $\mel{\psi_2}{\hat{v}_{geo}}{\psi_1}^{(2)}$. We can deduce the matrix element of $W^{(2)}_{21}$ following Eq.~\eqref{eq:psi_vgeo_psi_2} 
where $\mel{\psi_2^{(0)}}{\hat{v}_{geo}^{(1)}}{\psi_1^{(1)}} =\mel{\psi_2^{(1)}}{\hat{v}_{geo}^{(1)}}{\psi_1^{(0)}}=0$ due to the block-off diagonality of the matrix elements in the parallel-transport gauge. Therefore, we get a differential equation, Eq.~\eqref{eq:W2_mel_ODE}.

Here we adopt a verbose indexing notations, denoting the orbitals explicitly by band and momenta, $n_i\mathbf{k_i}$. 
To evaluate the first and second derivatives in Eq.~\eqref{eq:W2_mel_ODE} we first evaluate $\braket{\psi_{n_2\mathbf{k_2}}^{(0)}}{\psi_{n_1\mathbf{k_1}}^{(2)}}$ from the non-adiabatic second-order Sternheimer equation 
\begin{equation}\label{SI-eq:NA2_Stern}
\begin{split}
    &P_c(h_s^{(0)}-\epsilon_{n\mathbf{k}}^{(0)})P_c\ket{\psi_{n\mathbf{k}}^{(2)}}=\\
    &-P_c(\hat{v}_s^{(2)}+\hat{v}_{geo}^{(2)})\ket{\psi_{n\mathbf{k}}^{(0)}} -P_c(\hat{v}_s^{(1)}+\hat{v}_{geo}^{(1)})\ket{\psi_{n\mathbf{k}}^{(1)}}\;.
\end{split}
\end{equation}
For $n\mathbf{k} \in$ occupied subspace the consideration of the gauge choice eliminates the second term on the right hand side of the last equation. Projecting the equation on $\ket{\psi_{n_2\mathbf{k_2}}^{(0)}}$, we get 
\begin{equation}\label{SI-eq:braket2-0_1-2}
\begin{split}
    \braket{\psi_{n_2\mathbf{k_2}}^{(0)}}{\psi_{n_1\mathbf{k_1}}^{(2)}}
    =\frac{V^{(2)}_{n_2\mathbf{k_2},n_1\mathbf{k_1}}+W^{(2)}_{n_2\mathbf{k_2},n_1\mathbf{k_1}}}{\epsilon_{n_1\mathbf{k_1}}^{(0)}-\epsilon_{n_2\mathbf{k_2}}^{(0)}}\;,
\end{split}
\end{equation}
with $V^{(2)}_{n_2\mathbf{k_2},n_1\mathbf{k_1}}$, $W^{(2)}_{n_2\mathbf{k_2},n_1\mathbf{k_1}}$ defined above.
Substitution of Eq.~\eqref{SI-eq:braket2-0_1-2} in Eq.~\eqref{eq:W2_mel_ODE} leads to
the differential equation determining $W^{(2)}_{n_2\mathbf{k_2},n_1\mathbf{k_1}}$, Eq.~\eqref{eq:W2_mel_ODE_WV}.

In order to get an explicit form of this partial differential equation, one needs to work out the derivatives of $V_{n_2\mathbf{k_2},n_1\mathbf{k_1}}^{(2)}$.  
The $V_{n_2\mathbf{k_2},n_1\mathbf{k_1}}^{(2)}$ term, is given in Eq.~\eqref{eq:V2}. 
The derivative with respect to the phonon displacements $U_{\mathbf{q}\lambda}$ then follows

\begin{equation} \label{eq:1st_der_v2}
\begin{split}
    &\pdv{V^{(2)}_{n_2\mathbf{k_2},n_1\mathbf{k_1}}}{U_{\mathbf{q}\lambda}}= \sum_{\lambda'}\big[g^{(2)}_{n_2,n_1,\lambda,\lambda'}(\mathbf{k_2},\mathbf{q},\mathbf{q}') \frac{U_{\mathbf{q}'\lambda'}}{L_{\mathbf{q}'\lambda'}}\frac{1}{L_{\mathbf{q}\lambda}}\\
    &+g^{(2)}_{n_2,n_1,\lambda',\lambda}(\mathbf{k_1},\mathbf{q}',\mathbf{q}) \frac{U_{\mathbf{q}'\lambda'}}{L_{\mathbf{q}'\lambda'}}\frac{1}{L_{\mathbf{q}\lambda}}\big]\;,
\end{split}
\end{equation}
where $\mathbf{q}'=\mathbf{k}_2-\mathbf{k}_1-\mathbf{q}$.

We follow with a second derivation with respect to $U_{\mathbf{\bar{q}}\lambda}$, where $U_{\mathbf{\bar{q}}\lambda}=U_{\mathbf{q}\lambda}^*=U_{-\mathbf{q}\lambda}$. This requires  $\mathbf{k_2}=\mathbf{k_1}$. Hence
\begin{equation}\label{eq:2nd_der_v2}
\begin{split}
&\pdv{V^{(2)}_{n_2\mathbf{k_2},n_1\mathbf{k_1}}}{U_{\mathbf{\bar{q}}\lambda}}{U_{\mathbf{q}\lambda}}= \delta_{\mathbf{k_1},\mathbf{k_2}}  \frac{1}{L_{-\mathbf{q}\lambda}}\frac{1}{L_{\mathbf{q}\lambda}}\\
&\cross \left(g^{(2)}_{n_2,n_1,\lambda,\lambda}(\mathbf{k_1},\mathbf{q},\bar{\mathbf{q}})+g^{(2)}_{n_2,n_1,\lambda,\lambda}(\mathbf{k_1},\bar{\mathbf{q}},\mathbf{q})\right) \;.
\end{split}
\end{equation}
In light of these derived expressions, Eq.~\eqref{eq:1st_der_v2} and \eqref{eq:2nd_der_v2}, the differential equation in Eq.~\eqref{eq:W2_mel_ODE_WV} leads to Eq.~\eqref{eq:W2_ODE}.
The solution to the differential equation~\eqref{eq:W2_ODE}  
corresponds to the sum of a multiple of the homogeneous solution and the particular one. We distinguish two cases, one where $\mathbf{k_1}=\mathbf{k_2}$, leading to a second-order partial differential equation, and the other where $\mathbf{k_1}\neq \mathbf{k_2}$, amounting to a first-order differential equation. In equation  \eqref{eq:W2_sol} we represent particular solutions for the case $\mathbf{k_1}=\mathbf{k_2}$ and $\mathbf{k_1}\neq \mathbf{k_2}$ respectively. 

\section{}\label{app:B}
The transformation of the eigenenergies (Lagrange multipliers) from the parallel-transport gauge, (Eqs.~\eqref{eq-SI:orthonormal} and \eqref{eq-SI:paral_gauge}), to the diagonal gauge is done via a unitary transformation matrix \cite{gonze1995adiabatic} and is shown below.  
The eigenvalues in the parallel-transport gauge for the first- and second-order corrections are
\begin{equation}\label{SI-eq:e||1}
    \epsilon_{||,2,1}^{(1)} = \mel{\psi_2^{(0)}}{\hat{v}_s^{(1)}+\hat{v}_{geo}^{(1)}}{\psi_1^{(0)}} 
\end{equation}
and
\begin{equation}\label{SI-eq:e||2}
\begin{split}
    \epsilon_{||,2,1}^{(2)} &= 
    \mel{\psi_2^{(0)}}{\hat{v}_s^{(2)}+\hat{v}_{geo}^{(2)}}{\psi_1^{(0)}}\\
    &+ \mel{\psi_2^{(1)}}{\hat{v}_s^{(1)}+\hat{v}_{geo}^{(1)}}{\psi_1^{(0)}} 
    +\mel{\psi_2^{(0)}}{\hat{v}_s^{(1)}+\hat{v}_{geo}^{(1)}}{\psi_1^{(1)}} \\
    &+ \mel{\psi_2^{(1)}}{\hat{\mathcal{H}}_s^{(0)}}{\psi_1^{(1)}}
    -\frac{1}{2}\braket{\psi_2^{(1)}}{\psi_1^{(1)}}(\epsilon_2^{(0)}+\epsilon_1^{(0)})\;,
\end{split}
\end{equation}
respectively. 

The transformation from the parallel-transport gauge to the diagonal gauge for the eigenvalues follows:
\begin{equation} \label{SI-eq:ed2}
    \epsilon_{d,1}^{(2)}= \epsilon_{||,1,1}^{(2)} - \sum_{3}^{occ}\frac{\abs{\epsilon_{||,3,1}^{(1)}}^2}{\epsilon_3^{(0)}-\epsilon_1^{(0)}}\;,
\end{equation}
and by substituting Eq.~\eqref{SI-eq:e||1} and \eqref{SI-eq:e||2} it becomes
\begin{equation}\label{SI-eq:ed2_mel}
\begin{aligned}
    \epsilon_{d,1}^{(2)} = &\sum_{2\neq 1}^{occ}\frac{\abs{\mel{\psi^{(0)}_{2}}{\hat{v}_s^{(1)}+\hat{v}_{geo}^{(1)}}{\psi^{(0)}_{1}}}^2}{\epsilon^{(0)}_{1}-\epsilon^{(0)}_{2}}&&(a)\\
    &+ \mel{\psi_1^{(0)}}{\hat{v}_s^{(2)}+\hat{v}_{geo}^{(2)}}{\psi_1^{(0)}}&&(b)\\
    &+ \mel{\psi_1^{(1)}}{\hat{v}_s^{(1)}+\hat{v}_{geo}^{(1)}}{\psi_1^{(0)}}&&(c)\\
    &+ \mel{\psi_1^{(0)}}{\hat{v}_s^{(1)}+\hat{v}_{geo}^{(1)}}{\psi_1^{(1)}}&&(d)\\
    &+ \mel{\psi_1^{(1)}}{\hat{\mathcal{H}}_s^{(0)}}{\psi_1^{(1)}}&&(e)\\
    &-\frac{1}{2}\braket{\psi_1^{(1)}}{\psi_1^{(1)}}(\epsilon^{(0)}_{1}+\epsilon^{(0)}_{1})&&(f) \;.
\end{aligned}
\end{equation}

In order to get an analytical expression for $\epsilon_{d,1}^{(2)}$, we simplify the terms.
Using the definitions above for Eq.~\eqref{eq:1storder_wfn} in the parallel-transport gauge, the sum of terms $(c)-(f)$ becomes 
 \begin{equation}
     (c)+(d)+(e)+(f) =\sum_{2\neq 1}^{{unocc}}\frac{\abs{\mel{\psi^{(0)}_{1}}{\hat{v}_{s}^{(1)}+\hat{v}_{geo}^{(1)}}{\psi^{(0)}_{2}}}^2}{\epsilon^{(0)}_{1}-\epsilon^{(0)}_{2}}\;.
 \end{equation}

The second-order correction to the energies in the diagonal gauge follows  the addition of all the terms, resulting in Eq. \eqref{eq:en2_mel}.
The first term in Eq.~\eqref{eq:en2_mel} consists of first-order matrix elements that are given in
 Eq.~\eqref{eq:v_KS1},\eqref{eq:v_geo1}. The second term in Eq.~\eqref{eq:en2_mel} consists of the diagonal elements of the second-order matrix elements as in Eq.~\eqref{eq:V2} and Eq.~\eqref{eq:W2_mel_ODE}.
 Hence, the resulting term for the second-order correction to the eigenenergies in the diagonal gauge becomes 
\begin{equation}
    \begin{split}
       &\epsilon_{d,1}^{(2)} =\\
       & \sum_{2\neq 1} \sum_{\lambda\lambda'} \frac{g^*_{n_2,n_1,\lambda'}(\mathbf{k_1},\mathbf{q})\;g_{n_2,n_1,\lambda}(\mathbf{k_1},\mathbf{q})\; {(\epsilon^{(0)}_{1}}-\epsilon^{(0)}_{2})}{(\epsilon^{(0)}_{1}-\epsilon^{(0)}_{2}-\hbar\omega_{\mathbf{\bar{q}\lambda'}})(\epsilon^{(0)}_{1}-\epsilon^{(0)}_{2}-\hbar\omega_{\mathbf{\mathbf{q}\lambda}})} \frac{U_\mathbf{\mathbf{\bar{q}}\lambda'}}{L_{\mathbf{\mathbf{\bar{q}}\lambda'}}}\frac{U_\mathbf{\mathbf{q}\lambda}}{L_{\mathbf{\mathbf{q}\lambda}}}
    \\
       &+\sum_{\mathbf{q}\lambda} g^{(2)}_{n_1,n_1,\lambda,\lambda}(\mathbf{k}_1,\mathbf{q},\bar{\mathbf{q}}) \frac{\abs{U_{\mathbf{q}\lambda}}^2}{L_{\mathbf{q}\lambda}^2}\\
       &+ \sum_{2\neq 1}\sum_{\lambda\lambda'}\frac{g^*_{n_2,n_1,\lambda'}(\mathbf{k_1},\mathbf{q})g_{n_2,n_1,\lambda}(\mathbf{k_1},\mathbf{q})\hbar\omega_{\mathbf{q}\lambda}}{(\epsilon_1^{(0)}-\epsilon_2^{(0)}-\hbar\omega_{\mathbf{\bar{q}}\lambda'})(\epsilon_1^{(0)}-\epsilon_2^{(0)}-\hbar\omega_{\mathbf{q}\lambda})}\\ &\cross\bigg(\delta_{\lambda\lambda'} -\frac{U_{\mathbf{\bar{q}}\lambda'}}{L_{\mathbf{\bar{q}}\lambda'}}\frac{U_{\mathbf{q}\lambda}}{L_{\mathbf{q}\lambda}} \bigg)\\
       &-\sum_{2\neq 1}\sum_{\lambda\lambda'}\frac{g^*_{n_2,n_1,\lambda'}(\mathbf{k_1},\mathbf{q})g_{n_2,n_1,\lambda}(\mathbf{k_1},\mathbf{q})\hbar\omega_{\mathbf{q}\lambda}}{(\epsilon_1^{(0)}-\epsilon_2^{(0)}-\hbar\omega_{\mathbf{\bar{q}}\lambda'})(\epsilon_1^{(0)}-\epsilon_2^{(0)}-\hbar\omega_{\mathbf{q}\lambda})}\frac{U_{\mathbf{\bar{q}}\lambda'}}{L_{\mathbf{\bar{q}}\lambda'}}\frac{U_{\mathbf{q}\lambda}}{L_{\mathbf{q}\lambda}}\\
    \end{split}
\end{equation}
The sum of these leads to the more compact expression in Eq.~\eqref{eq:eps2_U}.

 \section{}\label{app:C}
Following the gauge transformation~\cite{gonze1995adiabatic} the second-order correction to the orbitals in the diagonal gauge is given in Eq.~\eqref{eq:psi||2_trans}.
The first term in Eq.~\eqref{eq:psi||2_trans} is the second-order correction to the wavefunction in the parallel-transport gauge, Eq.~\eqref{SI-eq:2_mel}.
The following terms consist of the unitary transformation matrix, $\mathcal{U}_{13}$, to first and second order in the perturbation,
\begin{equation}\label{SI-eq:U1}
    \mathcal{U}_{13}^{(1)*}=
    \begin{cases}
        0 & 1=3\\
        -\frac{\epsilon_{||,31}^{(1)}}{\epsilon_{3}^{(0)}-\epsilon_{1}^{(0)}} & 1\neq3
    \end{cases}
\end{equation}
\begin{equation}\label{SI-eq:U2}
\begin{split}
    \mathcal{U}_{13}^{(2)*}=
    \begin{cases}
        -\frac{1}{2}\sum_{2\neq3}^{occ}\abs{\mathcal{U}_{23}^{(1)}}^2 &1=3\\
        -\frac{1}{\epsilon^{(0)}_{3}-\epsilon^{(0)}_{1}}\bigg[\left(\sum_{2}^{occ}\mathcal{U}_{12}^{(1)*}\epsilon_{||,32}^{(1)}\right)\\
        \;\;\;\;\;\;+
        \epsilon_{||,31}^{(2)}-\epsilon_{d,1}^{(1)}\mathcal{U}_{13}^{(1)*}
        \bigg] &1\neq3 \;.
    \end{cases}
\end{split}
\end{equation}
These contain the first and second-order Lagrange multiplier, as given in Eq. \eqref{SI-eq:e||1} and Eq.~\eqref{SI-eq:e||2}, respectively.

We continue by evaluating the second term in Eq.~\eqref{eq:psi||2_trans} and substitute Eq.~\eqref{SI-eq:U1},\eqref{SI-eq:e||1},\eqref{eq:1storder_wfn} in the parallel gauge.
\begin{equation}\label{eq:U1term}
    \begin{split}
        \sum_3^{occ} \mathcal{U}_{13}^{(1)*}\ket{\psi^{(1)}_{3,||}}=
        &\sum_{3\neq1}^{occ} \left[-\frac{\epsilon_{||,31}^{(1)}}{\epsilon_{3}^{(0)}-\epsilon_{1}^{(0)}}\right] \ket{\psi^{(1)}_{3,||}}\\
        = &-\sum_{3\neq1}^{occ}\sum_{2}^{unocc} \frac{\mel{\psi^{(0)}_{2}}{\hat{v}_{s}^{(1)}+\hat{v}_{geo}^{(1)}}{\psi^{(0)}_{3}}}{(\epsilon_{3}^{(0)}-\epsilon_{1}^{(0)})}\\
        &\cross\frac{\mel{\psi_3^{(0)}}{\hat{v}_s^{(1)}+\hat{v}_{geo}^{(1)}}{\psi_1^{(0)}}}{(\epsilon^{(0)}_{3}-\epsilon^{(0)}_{2})}\ket{\psi^{(0)}_{2}}\;.
    \end{split}
\end{equation}

We then proceed to the third term in Eq.~\eqref{eq:psi||2_trans}, which consists of the second-order transformation matrix, Eq~\eqref{SI-eq:U2}. 
The $3=1$ term of the sum is
\begin{equation}\label{eq:U2term_diag}
    \begin{split}
        &\mathcal{U}_{11}^{(2)*}\ket{\psi^{(0)}_{1}}=
        \left[-\frac{1}{2}\sum_{4\neq1}^{occ}\abs{\mathcal{U}_{41}^{(1)}}^2\right]\ket{\psi^{(0)}_{1}}\\
        &=-\frac{1}{2}\sum_{4\neq1}^{occ}\abs{\frac{\mel{\psi_1^{(0)}}{\hat{v}_s^{(1)}+\hat{v}_{geo}^{(1)}}{\psi_4^{(0)}}}{\epsilon_{4}^{(0)}-\epsilon_{1}^{(0)}}}^2\ket{\psi^{(0)}_{1}}\;.
    \end{split}
\end{equation}
The off-diagonal term follows
\begin{equation}\label{eq:U2term}
    \begin{split}
        &\sum_{3\neq1}^{occ} \mathcal{U}_{13}^{(2)*}\ket{\psi^{(0)}_{3}}=\\
        &\sum_{3\neq1}^{occ} -\frac{\left(\sum_{4}^{occ}\mathcal{U}_{14}^{(1)*}\epsilon_{||,34}^{(1)}\right)
        +\epsilon_{||,31}^{(2)}-\epsilon_{d,1}^{(1)}\mathcal{U}_{13}^{(1)*}
        }{\epsilon^{(0)}_{3}-\epsilon^{(0)}_{1}}\ket{\psi^{(0)}_{3}}\;,
    \end{split}
\end{equation}
where, from Eq.~\eqref{SI-eq:U1} and~\eqref{SI-eq:e||1}, the first term in the numerator is 
\begin{equation}
    \begin{split}
    &\sum_{4}^{occ}\mathcal{U}_{14}^{(1)*}\epsilon_{||,34}^{(1)}=\\
    &\sum_{4}^{occ}\frac{\mel{\psi_4^{(0)}}{\hat{v}_s^{(1)}+\hat{v}_{geo}^{(1)}}{\psi_1^{(0)}}\mel{\psi_3^{(0)}}{\hat{v}_s^{(1)}+\hat{v}_{geo}^{(1)}}{\psi_4^{(0)}}}{(\epsilon_{4}^{(0)}-\epsilon_{1}^{(0)})}\;,
    \end{split}
\end{equation}
following Eq.~\eqref{SI-eq:e||2} the second term in the numerator is
\begin{equation}
\begin{split}
    &\epsilon_{||,3,1}^{(2)}= \mel{\psi_3^{(0)}}{\hat{v}_s^{(2)}+\hat{v}_{geo}^{(2)}}{\psi_1^{(0)}}\\
    &+ \frac{1}{2}\sum_{2}^{unocc} 
    \mel{\psi^{(0)}_{3}}{\hat{v}_{s}^{(1)}+\hat{v}_{geo}^{(1)}}{\psi^{(0)}_{2}} \mel{\psi^{(0)}_{2}}{\hat{v}_s^{(1)}+\hat{v}_{geo}^{(1)}}{\psi_1^{(0)}}\\
    &\cross\left(\frac{1}{(\epsilon^{(0)}_{3}-\epsilon^{(0)}_{2})}+\frac{1}{(\epsilon^{(0)}_{1}-\epsilon^{(0)}_{2})}\right)\;,
\end{split}
\end{equation}
and the energy correction in the third term in the numerator, $\epsilon_{d,1}^{(1)}$ vanishes.
\vspace{10pt}
Thus, Eq.~\eqref{eq:U2term} becomes
\begin{equation}\label{eq:U2term_final}
    \begin{split}
        &\sum_{3\neq1}^{occ} \mathcal{U}_{13}^{(2)*}\ket{\psi^{(0)}_{3}}=\\
        &\sum_{3\neq1}^{occ} \sum_{2}^{occ}\frac{\mel{\psi_3^{(0)}}{\hat{v}_s^{(1)}+\hat{v}_{geo}^{(1)}}{\psi_2^{(0)}}\mel{\psi_2^{(0)}}{\hat{v}_s^{(1)}+\hat{v}_{geo}^{(1)}}{\psi_1^{(0)}}}{(\epsilon_{3}^{(0)}-\epsilon_{1}^{(0)})(\epsilon_{2}^{(0)}-\epsilon_{1}^{(0)})} \ket{\psi^{(0)}_{3}}\\
        &-\sum_{3\neq1}^{occ}\frac{\mel{\psi_3^{(0)}}{\hat{v}_s^{(2)}+\hat{v}_{geo}^{(2)}}{\psi_1^{(0)}}}{\epsilon^{(0)}_{3}-\epsilon^{(0)}_{1}} \ket{\psi^{(0)}_{3}}\\
        &- \frac{1}{2}\sum_{3\neq1}^{occ}\sum_{2}^{unocc} 
        \frac{\mel{\psi^{(0)}_{3}}{\hat{v}_{s}^{(1)}+\hat{v}_{geo}^{(1)}}{\psi^{(0)}_{2}} \mel{\psi^{(0)}_{2}}{\hat{v}_s^{(1)}+\hat{v}_{geo}^{(1)}}{\psi_1^{(0)}}} {\epsilon^{(0)}_{3}-\epsilon^{(0)}_{1}}\\
        &\cross
        \left(\frac{1}{(\epsilon^{(0)}_{3}-\epsilon^{(0)}_{2})}+\frac{1}{(\epsilon^{(0)}_{1}-\epsilon^{(0)}_{2})}\right) \ket{\psi^{(0)}_{3}}\;.
    \end{split}
\end{equation}

Substituting into Eq.~\eqref{eq:psi||2_trans} the terms in Eq.~\eqref{SI-eq:2_mel},\eqref{eq:U1term},\eqref{eq:U2term_diag}, and~\eqref{eq:U2term_final},  leads to 
\begin{equation}
\begin{split}
    &\ket{\psi^{(2)}_{1,d}} =
    -\sum_{2\neq 1} \frac{V^{(2)}_{21}+W^{(2)}_{31}}{(\epsilon_{2}^{(0)}-\epsilon_{1}^{(0)})} \ket{\psi^{(0)}_{2}}
    -\frac{1}{2}\sum_{2\neq 1} \frac{\abs{V^{(1)}_{21}+W^{(1)}_{21}}^2}{(\epsilon_{2}^{(0)}-\epsilon_{1}^{(0)})^2} \ket{\psi^{(0)}_{1}}\\
    &-{\frac{1}{2}} \sum_{3\neq 1}^{occ}\sum_{2}^{unocc}  \frac{(V^{(1)}_{21}+W^{(1)}_{21})(V^{(1)}_{32}+V^{(1)}_{32})}{(\epsilon_{2}^{(0)}-\epsilon_{1}^{(0)})(\epsilon_{3}^{(0)}-\epsilon_{2}^{(0)})}\ket{\psi^{(0)}_{3}}\\
    &-\sum_{3\neq1}^{occ}\sum_{2}^{unocc} \frac{(V^{(1)}_{31}+W^{(1)}_{31}) (V^{(1)}_{32}+W^{(1)}_{32})}{(\epsilon_{3}^{(0)}-\epsilon_{1}^{(0)})(\epsilon^{(0)}_{3}-\epsilon^{(0)}_{2})}\ket{\psi^{(0)}_{2}}\\
    &+\sum_{3\neq1}^{occ} \sum_{2\neq1}^{occ}\frac{(V^{(1)}_{21}+W^{(1)}_{21})(V^{(1)}_{32}+W^{(1)}_{32})}{(\epsilon_{2}^{(0)}-\epsilon_{1}^{(0)})(\epsilon_{3}^{(0)}-\epsilon_{1}^{(0)})} \ket{\psi^{(0)}_{3}}\\
    &-\frac{1}{2}\sum_{3\neq1}^{occ}\sum_{2}^{unocc}  \frac{ (V^{(1)}_{21}+W^{(1)}_{21})(V^{(1)}_{32}+W^{(1)}_{32})} {\epsilon^{(0)}_{3}-\epsilon^{(0)}_{1}}\\
    &\cross\left(\frac{1}{(\epsilon^{(0)}_{3}-\epsilon^{(0)}_{2})}+\frac{1}{(\epsilon^{(0)}_{1}-\epsilon^{(0)}_{2})}\right) \ket{\psi^{(0)}_{3}}\;,
\end{split}
\end{equation}
with the definition $(V^{(1)}_{21}+W^{(1)}_{21})\equiv \mel{\psi^{(0)}_{2}}{\hat{v}_s^{(1)}+\hat{v}_{geo}^{(1)}}{\psi^{(0)}_{1}}$.
This corresponds to the second-order correction to the wavefunctions in the diagonal gauge. 
By plugging in the matrix elements of $\hat{v}_{s}$, Eq.~\eqref{eq:v_KS1}, \eqref{eq:V2}, and $\hat{v}_{geo}$, Eq.~\eqref{eq:v_geo1} and \eqref{eq:W2_sol}, one may get the explicit expression for the second-order non-adiabatic wavefunction correction in terms of DFPT components.

\end{document}